# Lognormal infection times of online information spread

Christian Doerr, Norbert Blenn, Piet Van Mieghem

Delft University of Technology, Delft, The Netherlands {C.Doerr, N.Blenn, P.F.A.VanMieghem}@tudelft.nl

**Abstract.** The infection times of individuals in online information spread such as the inter-arrival time of Twitter messages or the propagation time of news stories on a social media site can be explained through a convolution of lognormally distributed observation and reaction times of the individual participants. Experimental measurements support the lognormal shape of the individual contributing processes, and have resemblance to previously reported lognormal distributions of human behavior and contagious processes.

## 1 Introduction

The analysis of human social dynamics stemming from the emergent effects of individual human interactions has recently created a spur of research activity. Encompassing a wide area ranging from the propagation of opinions, epidemic spreading of information and innovation across groups of individuals [1], human dynamics investigates the temporal dependencies and interaction characteristics of the underlying contact and spreading processes, which are known to significantly drive overall diffusion and determine its contagion [2].

The analysis of the observed temporal distributions of online human activity data such as the inter-arrival and forwarding times of email [3], the propagation time of microblog posts [4], or telephony holding times [5], have commonly been approached and modeled [6] similar with the techniques established in topological network analysis, for example the approximation by power-law distributions. The applicability of fitting temporal behavioral data by a power-law has however been questioned [7,8] and bears a number of complications. First, the approximation through a power-law primarily concentrates on the fat tail providing a sub-optimal fit for the lower part of the observations, a power-law would imply that the highest frequency of data points is close to or equal to zero, yet our measurements show an up-going regime for those small values (see the example fit in fig. 1). Second, the relatively low power-law exponents found on temporal data militates against the presence of preferential attachment [9]. Third, while the process of preferential attachment provides an explanation for a scale-free degree distribution [10], this model does not provide insight into propagation time distributions. To this date there does not exist a theoretical model able to explain the observed traces of online human behavior. In this paper, we will put forward a working hypothesis to explain the propagation times generated via human behavior online.

Lognormal distributions [11] have frequently been associated with human behavior, such as the time to complete tasks, duration of strikes, income frequencies, epidemic incubation times or marriage age [12-14]. Additional, the cumulative behaviors of many individuals interacting with each other has also been demonstrated to result in a lognormal distribution [15], one theoretical explanation being given for example by the law of proportionate effect [16].

There are a few general processes that lead to a lognormal random variable. A lognormal random variable is defined [17] as $Y = e^X$, where $X$ is a Gaussian or normal random variable. The corresponding probability density function of the lognormal random variable $Y$ is

$$f_Y(t) = \frac{e^{-\frac{(\log t - \mu)^2}{2\sigma^2}}}{\sqrt{2\pi}\sigma t} \quad (1)$$

While a normal distribution describes well-behaved events, characterized by a mean and a standard deviation, such as the height of men and the temperature at noon in summer, the lognormal distribution exponentially blows up the relatively controlled deviations around the mean. Second, as a consequence of the Central Limit Theorem (CLT), the product of $n$ independent identically distributed (i.i.d.) random variables tends for large $n$ to a lognormal distribution. Slightly more general multiplicative processes (under certain conditions), such as the law of proportionate effect and geometric Brownian motion, give rise to lognormal behavior. Third, Marlow [18] beautifully demonstrates that, if a properly scaled sum of $n$ random variables tends to a normal distribution $N(0,1)$ for large $n$, the logarithm of the sum (also properly scaled) *also* tends to $N(0,1)$. In other words, both the scaled sum and the scaled logarithm of the sum converge to a same Gaussian $N(0,1)$, though at different rate. Yet differently stated, if the scaled sum of random variables tends to a Gaussian (by virtue of the CLT), then that same sum, though differently scaled, tends to a lognormal. Marlow's theorem proves the observation in radio communication that a sum of lognormal random variables also tends to a lognormal. These limit laws illustrate why lognormals may appear relatively frequently.

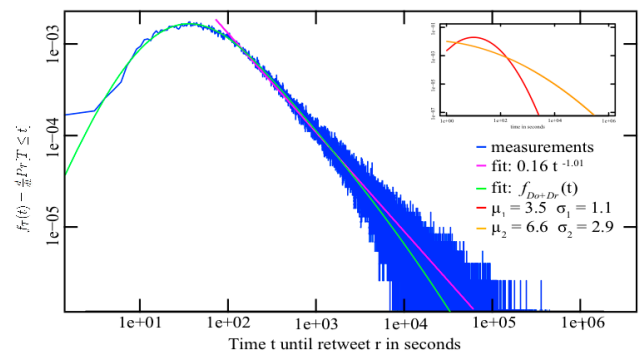

**Figure 1** The tail of the probability distribution of the inter-arrival time T of retweets on the microblog Twitter.com, here tweets reporting the death of Steve Jobs, can be approximated by a power-law, the entire distribution of tweets is however better fitted by a convolution of lognormals.

In this paper, we study the inter-arrival time of retweets in Twitter and the spreading times on Digg. Both times reflect human interaction through communication technology. We show that this relatively new type of human interaction is lognormally distributed.

## 2 The Distribution of the Spreading Time $T$

In order to spread information, such as forwarding a message or news item on a social media platform, three consecutive processes take place as shown in figure 2: First, after an e-mail, tweet or message has been sent by a person, the information is processed and physically copied across the network of servers of the large social media sites and is delivered into the inboxes and queues of the receiving users where they compete for the user's attention [19]. This first action requires $D_n$ time units, which we call the network propagation time. Second, users have to become aware of the content, for example by logging into the platform. The time between delivery and observation is denoted as the observation time $D_o$. Third, users decide to actively spread the information, for example through actions such as "retweets", "likes", "diggs", thereby effectively forwarding the message to their connected friends and/or followers. The time between observation and passing a message is the reaction time $D_r$. The overall person-to-person forwarding time $T$ is the sum of these three time components: $T = D_n + D_o + D_r$.

In the following discussion, time measurements from the microblog service Twitter and the (former) social news aggregator Digg.com are used to explore the spread of information on online social media. For the case of Twitter we measure $T$ as the time to forward ("retweet") messages by the followers of tweet originators, on Digg.com $T$ is denoted as the time between a user's recommendation (called a "digg") and the resulting diggs of a person's followers. For both these services, users see a summary of their friends' activities after visiting and logging into the website [20], and are presented with the opportunity to "retweet" or "digg" next to a particular piece of information.

Measurements of the network time $D_n$ indicated that $D_n$ is about two orders of magnitude smaller than the observation and reaction times. The network effect can therefore be neglected and the overall person-to-person forwarding time can be approximated by $T \approx D_r + D_o$. Hence, $T$ is, to a good approximation, determined by the habits and behaviors of the human participants.

In the following, we make two basic hypotheses:
- We assume that the random variable $D_r \geq 0$ is independent of $D_o \geq 0$, i.e., the time to react to a message does not depend on the observation time, which allows the probability density function of $T = D_o + D_r$ to be expressed as a convolution of those of $D_o$ and $D_r$:

$$f_{D_o+D_r}(t) = f_{D_o}(t) * f_{D_r}(t) \quad (2)$$
$$= \int_0^t f_{D_o}(x) f_{D_r}(t-x) dx \quad (3)$$

- We make the hypothesis based on previous findings [3, 13, 21], see discussion) that the observation and reaction times are lognormally distributed, $D_o \sim \text{logn}(\mu_o, \sigma_o)$ and $D_r \sim \text{logn}(\mu_r, \sigma_r)$, so that with (1),

$$f_{D_o+D_r}(t) = \frac{1}{2\pi\sigma_o\sigma_r} \int_0^t \frac{e^{-\frac{(log(t-x)-\mu_o)^2}{2\sigma_o^2}} e^{-\frac{(log(x)-\mu_r)^2}{2\sigma_r^2}}}{(t-x)x} dx \quad (4)$$

A maximum-likelihood estimation parameters of $(\mu_o, \sigma_o)$ and $(\mu_r, \sigma_r)$ for the measured spreading times on Twitter and Digg indeed generates a very good fit of the experimental data based on the two lognormally distributed time distributions. Figure 3 and 4 depict the experimental data and the maximum-likelihood fit of $f_{Do+Dr}(t)$ for the retweet time on Twitter and upvoting time on Digg respectively. The figure insets show the underlying lognormal distributions. The ML-fit is based on a log-squared error function which provides comparable attention to the entire distribution and does not overemphasize its tail, as discussed in [22]. The ML-fits are conducted using a set of 20.5 million tweets and 310 million diggs.

As the convolution is based on four variables, one would assume that a broad range of data could be fitted with a four-parameter function. We note that the convolution of a wide and narrow lognormal distribution is highly sensitive and only a combination of very small parameter ranges results in a good match with the observed Twitter data. Figure 5(a)-(d) depicts a mapping of all evaluated parameter combinations in the 4-dimensional fit optimization into four individual plots. The parameter combinations performing within 5% of the optimum are color-coded in yellow to red based on the goodness of fit and depicted in this color across all subfigures to be able to re-identify a particular scenario across the plots. As can be seen in figures (a) and (b), even small variations of $\mu_1$, $\sigma_1$ have a profound impact on the fit with a sub-exponential slope around the optimum, variations in $\mu_2$, $\sigma_2$ (figures (c) and (d)) have an attenuated but still significant impact as this distribution is very broad and mainly responsible for the tail of the convolution. As a result, only a few parameter combinations in a narrow region of each parameter space result in a good quality fit of the data. Moving away from the optimum in any of the four dimensions significantly degrades the goodness of the fit.

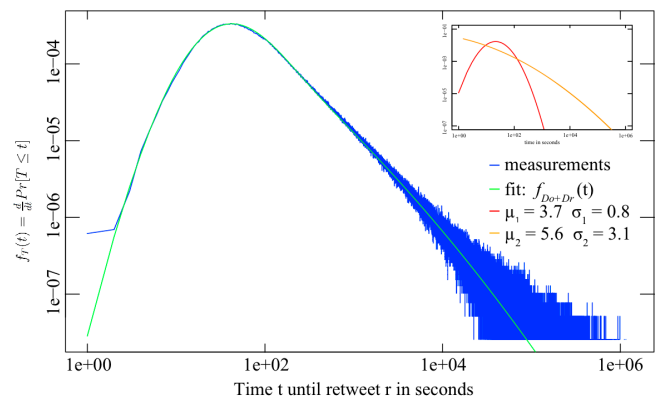

**Figure 3** Distribution of spreading times on Twitter.

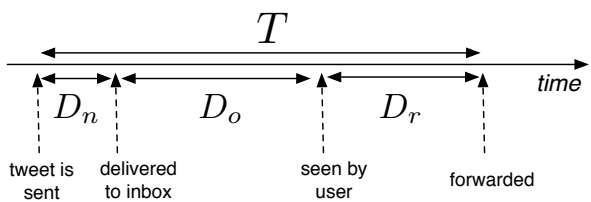

**Figure 2** The three components of network processing, user observation and user reaction time together form the measured inter-arrival time T.

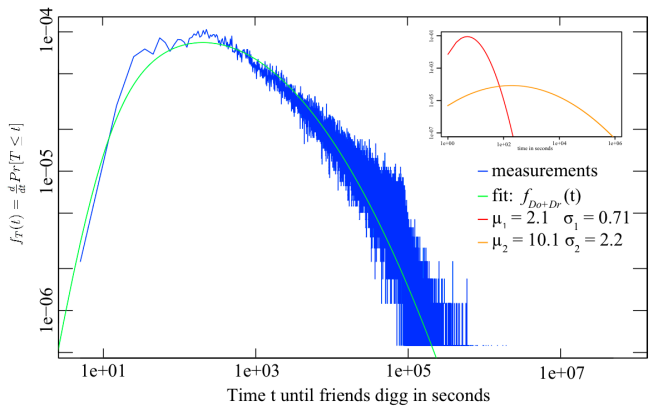

**Figure 4** Distribution of spreading times on Digg.

Similar patterns can be found across independent data measurements. Figure 1 shows analogously to figure 3 the distribution of retweet times for the spread of the news that the CEO of Apple, Steve Jobs, has died. Although this propagation is different from regular Twitter conversations shown in figure 3 in that most participants were likely triggered by external sources (traditional media coverage) and the high timely relevance of this news let the content spread an order of magnitude faster than for un-influenced retweets, the same process of two convoluted lognormal distributions fits the observed data well.

The accuracy and the high sensitivity of the fit leads us to conclude that the two assumptions (independence and lognormal distribution of the time components $D_o$ and $D_r$) are realistic.

## 3 Identifying Observation and Reaction Time

As a convolution of two functions is commutative, $f * g = g * f$, it is not possible at this point to identify which of the two lognormal distributions describes observation and reaction time respectively. This section will therefore use empirical data from Digg.com and Twitter to make this differentiation. As directly obtaining measurements to assess the observation and reaction time is not feasible, these times will be indirectly inferred by monitoring a particular web site, and determine from the publicly visible record of status updates, friendship formations or comments when a person is active or not.

For the differentiation in the case of Digg.com, we utilize data on the times that their registered users upvoted or commented on a story published within the social news aggregator that until the sale of Digg.com in July 2012 could be publicly downloaded via its API. From this timestamped record of 310 million individual upvotes, we reconstruct the approximate coherent time periods a particular user is active on the website and thereby may spread information (figure 6). As the epidemic spread of information assumes the diffusion along

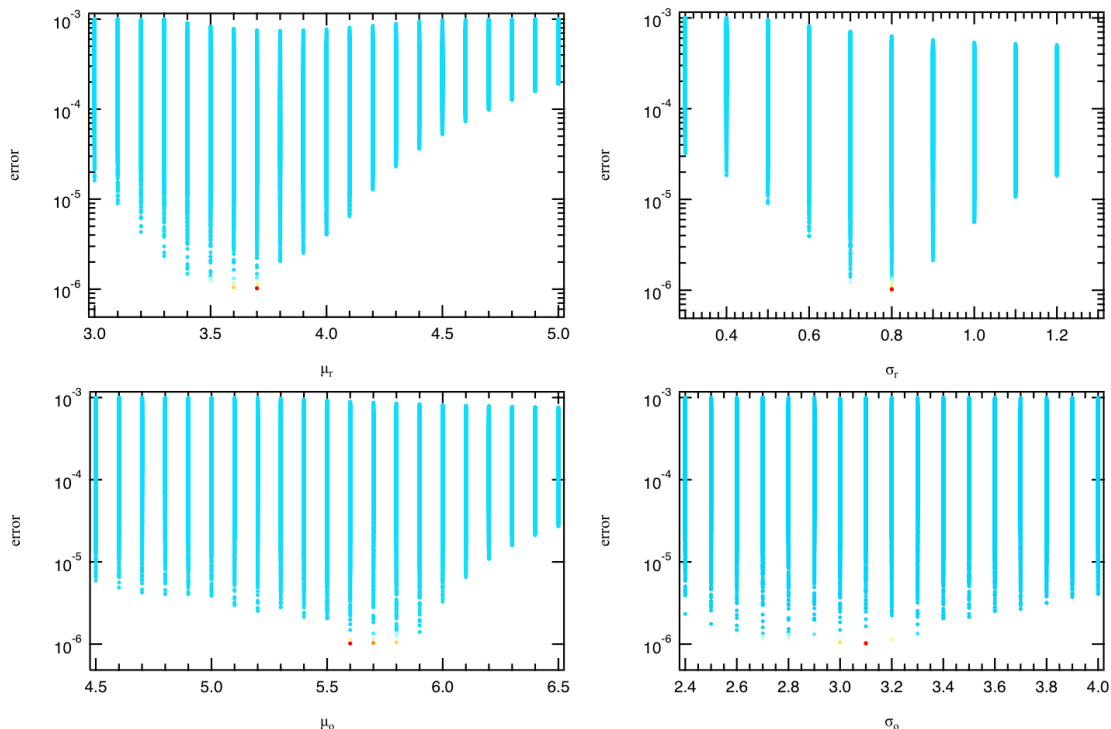

**Figure 5** Error of the lognormal fit for Twitter of Figure 3.

     

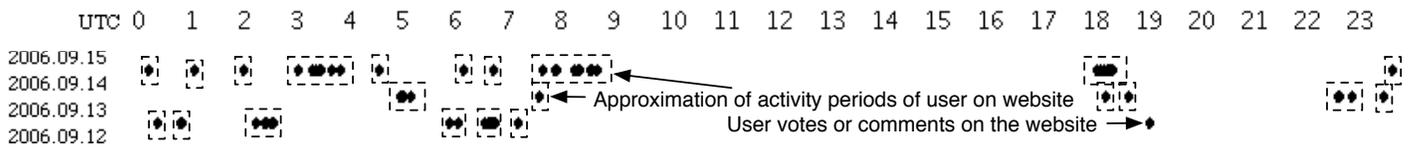

**Figure 6** The record of a user's votes on the Digg.com website, each indicated by a dot on the timeline, provides a lower bound estimation of the user's presence on the website, as well as an upper bound on this user's observation times, exemplary for four consecutive days in Sep. 2006.

social relationships, we augment this simple view through information about the unidirectional follower and bi-directional friendship relations (i.e., reciprocal following) that users are forming on the Digg network. When "following" a user, the follower receives notifications about the other person's activities with the possibility to upvote and thereby spread the information in turn to his own followers and friends. The observation time in an epidemic spreading process across these social relationships is the time between a particular user has upvoted a news item, and this recommendation will be displayed on the follower's user interface on his next login on the website. The reaction time equals the interval between the notification and the receiver's upvote (if applicable).

Since the public record of diggs and comments does not directly list the instances at which a user has consulted his inbox to see the incoming notifications during a particular visit on the website, this monitoring procedure will only create a general bound of the observation and reaction times (see figure 7). As the notification page is only visible after a login (and the login event must have been completed by the time we observed the first digg), the time between the original digg and the approximated login time of the follower will be lower bound of the observation time, provided that a particular person could have checked the incoming recommendations at any time point after the login. Correspondingly, the time between the login event and the person's own digg will therefore be *an upper bound of the reaction time*, as the notification triggering the digg could have been read anytime between the login and the actual action. The same approach is followed in case of Twitter, where the timestamps of tweets will be used to establish an estimate when a user is present on the social network, leading upper and lower bounds of reaction and observation time respectively.

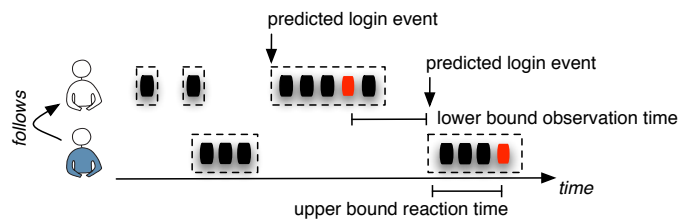

**Figure 7** A lower bound estimate of the observation time and an upper bound of the user's reaction times can be determined when augmenting the voting behavior with the friendship network.

The analysis of the experimental data shows that both lower and upper bound of observation and reaction time are following a lognormal distribution, thereby confirming our initial hypotheses for the spreading time *T*. Table 1 list the ML-fit based on (4) against the experimentally obtained parameters of the underlying distributions. The two distributions from the maximum-likelihood fits also line up on the correct side and neighborhood of the empirical lower and upper bounds, thereby making it possible to *identify $\mu_1$, $\sigma_1$ as the reaction time (colored red in the insets of figures 1, 3 and 4), and $\mu_2$, $\sigma_2$ as the observation time (colored orange in the insets)*. The measured and fitted observation time parameters $\mu_o$, $\sigma_o$ match well those predicted from our earlier maximum-likelihood fit. The mean of the measured reaction time also falls within the expected interval, between 0 and the computed upper bound, although the bound is less tight.

## 4 Materials and Methods

The analysis of spreading times was based on two datasets, a collection of initial original posts and their corresponding retweets on the microblog Twitter, and five year trace of stories and votes submitted to the social media aggregator Digg.com. Both datasets were collected from the providers' webpages and API interfaces and contain only publicly available information. The former dataset of 20.5 million tweets was constructed in a two step process: Starting from an initial 1% random sample of all tweets published within Twitter, we automatically retrieved for each of these initial tweets all further subsequent reappearances of these messages thereby generating forests of individual message spreads [19]. For all persons involved in the propagation of these messages, a complete trace of their social relations was retrieved. The latter dataset is comprised of the content submissions to the social media aggregator Digg.com within the seven year period of 2004-2010 and the recommending votes of its users. Using the crawling technique described in [20], we obtained a complete record of the 310 million votes from 2.2 million users' activity on the 12 million submitted stories.

## 5 Discussion

The analysis and the experimental data point to lognormal distributions as an explanation of the temporal distributions observed in epidemic spreading online. This finding is not per se surprising as lognormal distributions have for a long time been observed in and connected to human behavior, which is also driving the spread of information and innovation online.

The analogue to the reaction time $D_r$ is, for example, in classical epidemiology referred to as the incubation period. This incubation time in epidemic processes between the infection of an individual and the first symptoms or active transmission of the disease is frequently found to adhere to a lognormal distribution, for example in the case of chicken pox, hepatitis, or salmonellosis [12, 23]. In an overview of incubation periods of 86 diseases, Nishiura concludes that 70.9% can be accepted as lognormal at a 5% level of significance [21].

**Table 1** Comparison of observation and reaction times of ML-fit and derived bounds from Digg and Twitter dataset.

| Measurement | Digg | | Twitter | |
|---|---|---|---|---|
| | μ | σ | μ | σ |
| Lognormal $D_o$ (ML-Fit) | 10.2 | 2.3 | 5.6 | 3.1 |
| Lower Bound of Observation Time | 9.48 | 2.75 | 5.5 | 2.9 |
| Lognormal $D_r$ (ML-Fit) | 2.1 | 0.71 | 3.7 | 0.8 |
| Upper Bound of Reaction Time | 6.26 | 2.02 | 6.92 | 1.32 |

Similar lognormal patterns have been discovered across domains for the equivalent of the observation time $D_o$. Barcelo [24] reports the channel holding time on cellular networks, the duration of calls and thus the time of the transmission process to be lognormally distributed.

This general notion of the lognormally distributed duration of human activities and associated spreading tasks are also repeatedly found across other domains, for example in the duration of strikes [14] or the time to complete tasks on a test [13]. This same pattern seems to extend into the online domain for spreading times, for example in the response times of viral marketing campaigns as reported by Iribarren and Moro [3].

The lognormal distribution regularly appears in the broader context of universal human activities and behavior. When tracing the mobility of cell phone handsets, Barcelo and Jordan [5] find that the time a person stays at a certain location (and is associated with a particular cell phone cell) also follows a lognormal distribution, which at near 100% penetration in many countries can be seen as a good proxy to measure human mobility. Radicchi, Fortunato and Castellano [25] for example report that the citation counts of academic articles universally follows a lognormal distribution when normalized by the relative citation habits in a scientific field. A similar phenomenon can also be observed in the law of proportionate effect [16], demonstrating that the publicly visible total number of recommendations changes the voting behavior of individuals in Digg and leads at a population-level to results best characterized by the lognormal distribution [16].

Similarly, a variety of such lognormally distributed properties of populations have been discovered, such as in the distribution of incomes or marriage ages [12]. Shockley [26] explains the lognormally distributed productivity of researchers, which seems related to the observed, universal lognormal citation counts of Radicchi *et al.* [25].

# 6 Conclusion

Based on a huge number of measurements in Twitter and Digg, we found that human related activities such as the observation time $D_o$ and reaction time $D_r$, are very close to a lognormal random variable. While favoring power-law behavior, Vazquez et al. [27] argue that a proper identification of a lognormal (and its distinction from the ubiquitously present power-law) distributions requires the availability of extensive datasets. With an analysis of two datasets of 310 and 20.5 million interactions, we believe that the presented verification here amply confirms the lognormal behavior. Although there is theoretical support for the occurrence of lognormal random variables, the precise underlying stochastic processes, that lead to a lognormal outcome are today still missing.

# Acknowledgments

This research has been supported by the EU Network of Excellence in Internet Science EINS (project no. 288021) and the Trans-sector Research Academy for Complex Networks and Services (TRANS).